\documentstyle[amssymb,12pt,graphics,aps]{revtex}
\begin{document}
\draft
\title{Statistical Properties of Quantum Graph Spectra}
\author{Yu. Dabaghian}
\address{Department of Physiology,
Keck Center for Integrative Neuroscience,\\
University of California,
San Francisco, California 94143-0444, USA \\
e-mail yura@phy.ucsf.edu}
\date{\today}
\maketitle

\begin{abstract}
A general analytical approach to the statistical description of quantum 
graph spectra based on the exact periodic orbit expansions of quantum levels 
is discussed. The exact and approximate expressions obtained in \cite{Anima} 
for the probability distribution functions using the spectral hierarchy method 
are analyzed. In addition, the mechanism of appearance of the universal 
statistical properties of spectral fluctuations of quantum-chaotic systems 
is considered in terms of the semiclassical theory of periodic orbits.
\end{abstract}


\newpage

\section{Introduction}

A quantum graph system consists of a quantum particle moving along the bonds 
of an arbitrary finite graph $G$ \cite{QGT}. In the classical limit, this 
system generates a simple stochastic dynamics, which is specified by the 
translational motion along the bonds of the graph and stochastic scattering 
at its vertices with preset scattering probabilities. This dynamics has many 
common features with the dynamics of usual chaotic systems \cite{Gaspard}.
For example, periodic trajectories in such a system are isolated and their 
number increases exponentially with the period. At the same time, the statistical 
behavior of various spectral characteristics of sufficiently complex quantum 
graphs, e.g. the probability distribution of spacings $s_{n}=k_{n}-k_{n-1}$ 
between the nearest levels of the momentum was numerically shown \cite{QGT} 
to follow the predictions of the Random Matrix Theory (RMT) \cite{BGS,Zaslavsky}, 
as it is usually the case for classically nonintegrable systems. It also turns 
out that a great number of problems of classical and quantum dynamics on the 
graph allow exact solutions, which makes these systems convenient models in 
the context of the analytical theory of ``quantum chaos''. In particular, for 
these systems there exist the exact periodic orbit expansions of the quantum 
density of states (Gutzwiller formula) \cite{QGT} along with a similar expansion 
for the spectral staircase:
\begin{equation}
N(k)\equiv \sum_{j=1}^{\infty }\Theta \left( k-k_{j}\right)=\bar{N}(k)+
\frac{1}{\pi}\mathop{\rm Im}\sum_{p}A_{p}e^{iL_{p}^{(0)}k},
\label{gutzw}
\end{equation}
Here $\bar{N}(k)$ is the average number of levels in the range $\left[0,k\right]$, 
$L_{p}^{(0)}$ is the optical length of the periodic trajectory with the index $p$, 
and $A_{p}$ is a certain weight factor explicitly defined in terms of the scattering 
coefficients at the graph vertices.
It should be emphasized that the existence of the explicit expansions of the global 
characteristics such as (\ref{gutzw}) is not equivalent to the ultimate solution of 
the spectral problem, which should provide local information about the individual 
levels in the form of an explicit dependence $k_{n}=k(n)$. An approach for determining 
the quantities $k_{n}$ explicitly was proposed in \cite{Anima}, which is based on 
using a finite system of $r+2$ auxiliary ``separators'' $\hat{k}_{n}^{(0)}$, 
$\hat{k}_{n+1}^{(1)}$, ..., $\hat{k}_{n}^{(r+1) }$,  the first of which is the physical 
spectral sequence $k_{n}=\hat{k}_{n}^{(0)}$, and the last one is a globally defined 
explicit function of $n$:
\begin{equation}
\hat{k}_{n}^{(r+1) }=\frac{\pi }{L_{0}}\left( n+\frac{1}{2}\right).
\label{mid}
\end{equation}
The key property of these sequences is that they must satisfy the ``bootstrapping'' 
conditions
\begin{equation}
\hat{k}_{n}^{(j)}<\hat{k}_{n}^{(j-1)}<\hat{k}_{n+1}^{(j)},
\label{bootstrap}
\end{equation}
which guarantee that between every pair of the neighboring points $\hat{k}_{n}^{(j)}$ 
and $\hat{k}_{n+1}^{(j)}$ (see Fig. 1) there exists  a single point $\hat{k}_{n}^{(j-1)}$.
In \cite{Anima} it was also pointed out that due to certain analytical properties of the 
spectral determinant $\Delta (k)=1+\sum_{i}a_{i}e^{ikL_{(i)}}$, where $L_{(i)}$, are 
different linear combinations for the bond lengths $l_{1},l_{2}$, ..., $l_{N_{B}}$, the 
set $\hat{k}_{n}^{(j)}$ can be provided by the sequence of zeros of the $j$-th derivative 
of the function $\Delta (k)$ \cite{Anima,LevinBY}. In this case, the quantity $r$ characterizing 
the degree of spectral irregularity is defined as the minimal number for which the condition 
$\sum_{i}\left\vert a_{i}\left(L_{(i)}/L_{0}\right) ^{r}\right\vert <1$ is satisfied \cite{Anima}. 
\begin{figure}
\begin{center}
\includegraphics{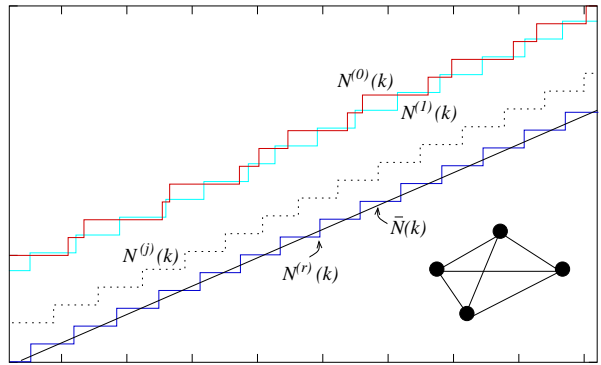}
\end{center}
\caption{Bootstrapping of spectral staircases for separating
sequences $\hat{k}_{n}^{(j)}$ of the completely connected four-vertex
graph with $r=7$. The plots $N^{(j)}(k)$  are vertically shifted for
the sake of clarity. It is clear that the physical spectral staircase 
$N^{(0)}(k)$  is interlaced by the staircase $N^{(1)}(k)$, etc. The last 
staircase $N^{(r)}(k)$  is intersected by the Weyl average $\bar{N}(k)$}
\label{Fig.1} 
\end{figure}
In the simplest case of regular graphs when $r = 0$ \cite{Opus,Prima}, only one auxiliary 
sequence (\ref{mid}) is required and various spectral characteristics can be calculated using 
the formula
\begin{equation}
f\left( k_{n}\right) =\int_{\hat{k}_{n-1}^{(1)}}^{\hat{k}
_{n}^{(1)}}f(k)\rho (k)dk.
\label{leveldelta}
\end{equation}
As pointed out in \cite{Opus,Anima}, this case corresponds to the situation in which the 
straight line with the slope $L_{0}/\pi $,  representing the Weyl average $\bar{N}(k)$,
``pierces'' the physical spectral staircase $N(k)$,  i.e., $\bar{N}(k)$  intersects every 
stair step of $N(k)$  at the points $\hat{k}_{n}^{(1)}$.

\section{Statistical properties of the spectra of regular graphs}

Using the Gutzwiller formula in Eq. (\ref{leveldelta}), one can derive the explicit 
expansions for various spectral characteristics $f_{n}^{(0)}$, for example, for fluctuations 
$\delta _{n}^{(0)}=\frac{L_{0}}{\pi }\left(k_{n}-\bar{k}_{n}\right) $, of the eigenvalues 
$k_{n}$ around the Weyl average or for the distances between levels $s_{n,m}=k_{n+m}-k_{n}$. 
Such expansions have the form \cite{Opus,Prima}
\begin{equation}
f_{n}^{(0)}=\bar{f}^{(0)}-\sum_{p}C_{p}^{(0)}\,\cos \left(\omega _{p}^{(0)}n+\varphi _{p}^{(0)}\right) ,  
\label{fn}
\end{equation}
where the frequencies $\omega _{p}^{(0)}$ are defined via the periodic orbit lengths as, 
$\omega _{p}^{(0)}=\pi L_{p}^{(0)}/L_{0}$. The first term of expansion (\ref{fn}) determines 
the average value of the quantity $f_{n}^{(0)}$, whereas the following sum describes 
fluctuations around the average. Each frequency $\omega_{p}^{(0)}$ is an integer combination 
$\omega _{p}^{(0)}=m_{p,1}^{(0)}\Omega_{1}
+m_{p,2}^{(0)}\Omega_{2}+...+m_{p,N_{B}}^{(0)}\Omega_{N_{B}}$, of the quantities $\Omega_{i}$, 
which are expressed in terms of the lengths of the graph bonds as $\Omega _{i}=l_{i}/L_{0}$, 
and the coefficients $m_{p,i}^{(0)} $ indicate how many times the orbit passes along the bond 
$l_{i}$. 
The sum $\left\vert m_{p}^{(0)}\right\vert =m_{p,1}^{(0)} +m_{p,2}^{(0)}+...+m_{p,N_{B}-1}^{(0)}$ 
specifies the total number of scattering events that the particle moving along the trajectory $p$ 
undergoes at the vertices. If Eq. (\ref{fn}) includes only the orbits for which $|m_{p}^{(0)}|<m$, 
we arrive at the $m$-th approximation to the exact value $f_{n}^{(0)}$ \cite{QGT,Opus}.

Since the numbers $\Omega _{i}$ satisfy the condition $\Omega _{1}+\Omega_{2}+...+\Omega _{N_{B}}=1$, 
only $N_{B}-1$ of these numbers are independent. Expressing one of them, e.g., $\Omega _{N_{B}}$, in 
terms of the others, let us consider the (generic) case when the numbers 
$\tilde{\Omega} _{i}=\Omega _{i}-\Omega _{N_{B}}$ are irrational and algebraically independent. Let 
us call the orbit $p$ algebraically simple (with the notation $p'$) if the integer coefficients 
$\tilde{m}_{p,i}^{(0)}=m_{p,i}^{(0)}-m_{p,N_{B}}^{(0)}$ have no common divisors. Such orbits in general 
differ from the {\em dynamically} simple orbits that correspond to single traversals along closed 
sequences of bonds during the particle's motion along the graph \cite{QGT,Gaspard,Anima,Opus,Prima}.

The expansion (\ref{fn}) enables one to pass immediately to the statistical description of the sequence 
$f_{n}^{(0)}$. Indeed, it is well known that the sequence of the remainders 
$x_{n}=\left[\alpha n\right]_{\mathop{\rm mod}1}$ for any irrational number $\alpha $ and $n=1,2,...$, 
is uniformly distributed in the interval $\left[ 0,1\right] $ \cite{Karatsuba}. Since the arguments of 
the trigonometric functions appearing in series (\ref{fn})  are defined modulo $2\pi $, parsing through 
the values $f_{n}^{(0)}$ yields a sequence which is statistically equivalent to the series
\begin{equation}
f_{x}^{(0)}=\bar{f}^{(0)}-\sum_{p}\tilde{C}_{p}^{(0)}\,\sin
\left( \tilde{m}_{p}^{(0)}x+\varphi _{p}^{(0)}\right),
\label{fxind}
\end{equation}
Here, $\tilde{C}_{p}^{(0)}$ and $\tilde{m}_{p}^{(0)}$ correspond to the coefficients of Eq. (\ref{fn}) 
in which the condition $\sum_{i}\Omega_i=1$ is taken into account, and $x$ is a set of $N_{B}-1$ 
independent, uniformly distributed random variables. The distribution of the quantities $\delta f_{x}^{(0)}$ 
in this case is obtained from the expression $P_{f}^{(0)} =\langle\delta \left(f^{(0)}-f_{x}^{(0)}\right)\rangle$:
\begin{equation}
P_{f}^{(0)} =\int dke^{ik\left( f^{(0)}-\bar{f}^{(0)}\right)}
\int_{0}^{2\pi}\prod_{p}\Lambda_{p}(x)\frac{dx}{2\pi } 
\label{pf}
\end{equation}
where every factor $\Lambda _{p}( \vec{x}) =e^{ik\tilde{C}_{p}^{(0)}\,\cos
\left( \tilde{m}_{p}^{(0)}x+\varphi _{p}^{(0)}\right) }$
determines the contribution to the integral from the corresponding periodic orbit $p$. Thus, Eq. (\ref{pf}) 
gives the exact expression for the distribution $P_{f}^{(0)}$ in terms of the periodic orbit theory.
It is important to point out that the properties of the asymptotic distributions of trigonometric sums 
of form (\ref{fxind}) are one of the traditional areas of research of mathematical statistics (see, e.g., 
\cite{Proxorov,Revesz} and references therein). In particular, it is known that separate terms (or groups 
of terms) of lacunary trigonometric series of form (\ref{fxind})  can be considered as weakly dependent 
random variables, for which one can be establish a generalization of the central limit theorem, and 
consequently their sum is asymptotically Gauss distributed according to
\begin{equation}
P_{f}^{(0)} =\frac{1}{\sigma \sqrt{
2\pi }}e^{-\frac{\left( \delta f^{(0)}\right) ^{2}}{2\sigma ^{2}}},
\label{gauss}
\end{equation}
with the variance
\begin{equation}
\sigma ^{2}=\frac{1}{2}\sum_{p}\tilde{C}_{p}^{(0)2}=\left\langle
\left( \delta f_{x}^{(0)}\right) ^{2}\right\rangle.
\label{variance}
\end{equation}
The conclusion about the Gaussian form of the distribution of the fluctuations also appears to be 
applicable to this kind of spectral characteristics expansions of most of the regular quantum graphs (and 
other scaling systems), which are described by series of form (\ref{fxind})  with constant coefficients.
A hypothesis about the Gaussian nature of the distribution the spectral staircase fluctuations 
$\delta N(k)=N(k)-\bar{N}(k)$, confirmed by extensive numerical investigations, was previously 
proposed in \cite{ABS} as the universal ``central limit theorem for spectral fluctuations''
applicable to general quantum chaotic systems. Owing to the existence of additional explicit 
expansions (\ref{fn}) this hypothesis, corroborated by the relation with the theory of weakly 
dependent random variables (trigonometric sums) can actually be extended to a much wider set of 
spectral characteristics.

\section{Approximate description of the distribution functions}
Since the contributions of individual orbits to the series $\delta f_{x}^{(0)}$  behave as 
weakly dependent random variables, some physical simplifications are possible in Eq. (\ref{pf}). 
Expanding the exponentials $\Lambda _{p}( \vec{x})$, one can note that because expansion (\ref{fn}) 
is made in orthogonal harmonics, most integrals of the cross terms appearing from the product of 
the expansions $\Lambda _{p}(\vec{x})$ in Eq. (\ref{pf}). Contributions come only from the ``resonant'' 
terms for which one of the algebraic sums of the frequencies vanishes. The amplitude of these 
contributions decreases rapidly in the orders of the corresponding degrees of $C^{(0)}_{p}$, 
that are proportional to the product of the corresponding number of scattering coefficients at 
graph vertices \cite{Opus,Prima}.
This argumentation can be used to simplify the integral for $P_{f}^{(0)}$. For example, in a 
simple approximation the contributions from resonances between different algebraically simple 
orbits can be disregarded. This is equivalent to untangling of the factors $\Lambda _{p'}(x)$  
corresponding to different algebraically simple orbits, i.e., to the introducing an independent 
set of variables $x_{p'}$ for every algebraically simple orbit. In this case, the distribution 
probability is represented in the form
\begin{equation}
P_{f}^{(0)} =\int dke^{ik\left( f^{(0)}-
\bar{f}^{(0)}\right) }\prod_{p'}Q_{p'}\left(k\tilde{C}^{(0)}_{p'}\right) ,  
\label{prime}
\end{equation}
where every factor
\begin{equation}
Q_{p'}=\int_{0}^{2\pi}e^{ik\sum_{\nu }\tilde{C}_{p'\nu}^{(0)}\,
\cos \left( \nu \tilde{m}_{p'}x_{p'}+\varphi_{p}\right) }dx_{p'},  
\label{primep}
\end{equation}
corresponds to the algebraically simple orbit $p'$ and the sum with respect to $\nu $ 
in Eq. (\ref{primep}) is calculated over orbits whose indices are multiples of $\tilde{m}_{p'}$.
For a more crude description of the probability distribution profile, one can disregard the resonances 
between any distinct orbits, which is equivalent to the introduction of an independent phase $x_{p}$ 
for every orbit. Under this assumption, the integral in Eq. (\ref{pf}) is separated into independent 
integrals and, as a result, we arrive at the simple expression
\begin{equation}
P_{f}^{(0)} =\int dke^{ik\left( f^{(0)}-\bar{f}^{(0)}\right)}
\prod_{p}J_{0}\left( k\tilde{C}_{p}^{(0)}\right) ,
\label{distreg}
\end{equation}
where $J_{0}(x)$  is the zeroth Bessel function.
Distributions of form (\ref{distreg}) appear in communication theory, for example, when analyzing the 
intensity of interfering telecommunication channels, the theory of wave propagation in random media, 
and other fields where stochastic signal models are used \cite{LevinBR,Paetzold}.
It is also worth noting that, in the approximation of independent random contributions, the conditions 
of the Lindeberg--Feller theorem and central limit theorem are satisfied, which establish the normal 
distribution law for the sum of independent random variables. For spectral expansions (\ref{fxind}) 
these conditions on the variances $\sigma_{p}^{2}=\left(\tilde{C}_{p}^{(0)}\right) ^{2}/2$ of individual 
contributions are satisfied due to the exponential increase in the number of periodic orbits and the 
uniform exponential decrease of the magnitude of the coefficients $\tilde{C}_{p}^{(0)}$. As a result, 
in the approximation of independent random contributions, distribution (\ref{distreg}) has the same 
Gaussian form (\ref{gauss}), with the variance $\sigma ^{2}=\sum_{p}\tilde{C}_{p}^{(0)2}/2<\infty$ as 
that predicted in \cite{Proxorov,Revesz} and \cite{ABS} for the case of weakly dependent variables.
Such description is applicable to the statistical properties of various spectral characteristics of
the regular graphs beginning with their harmonic expansions \cite{Anima,Opus,Prima}. For example one 
can consider the fluctuations $\delta _{n}^{(0)}=\frac{L_{0}}{\pi}\left( k_{n}-\bar{k}_{n}\right) $, 
of levels around the average value, which have form (\ref{fn}) with $\bar{\delta}^{(0)}=0$, 
$\varphi _{p}^{(0)}=-\frac{\pi }{2}$, and the coefficients
\begin{equation}
C_{p}^{(0)}=-\frac{2}{\pi }\frac{A_{p}^{(0)}}{\omega _{p}}\sin \left( \frac{
\omega _{p}}{2}\right) ,
\label{cp}
\end{equation}
or the difference $s_{m,n}^{(0)}=k_{n+m}-k_{n}$ with $\bar{s}_{m,n}^{(0)}=$ $\frac{\pi }{
L_{0}}m$, $\varphi _{p}^{(0)}=\frac{\omega _{p}m}{2}$ and the coefficients
\begin{equation}
D_{p,m}^{(0)}=\frac{4}{L_{0}}\frac{A_{p}^{(0)}}{\omega _{p}}\sin \left( 
\frac{\omega _{p}}{2}\right) \sin \left( \frac{\omega _{p}m}{2}\right) .
\label{dp}
\end{equation}
Knowing the distributions of these quantities, one can describe more complex objects such as the 
correlation function of fluctuations $\left\langle \delta_{n}^{(0)}\delta_{n+m}^{(0)}\right\rangle$, 
autocorrelation function $R_{2}(x)$, and the form factor $K_{2}(\tau)$, given by the expression
\begin{equation}
K_{2}=\frac{\pi }{L_{0}}\sum_{m}\left\langle e^{-is_{mn}\tau}\right\rangle=
\frac{\pi }{L_{0}}\sum_{m}e^{-i\frac{\pi m}{L_{0}}\tau }F_{s_{m}}^{(0)}(k),
\label{k2}
\end{equation}
where $F_{s_{m}}^{(0)}(k)$  is the characteristic function of distributions of form (\ref{pf}), 
(\ref{prime}) or (\ref{distreg}), which are obtained from expansion (\ref{fn}) for $s_{m,n}$  with 
coefficients (\ref{dp}), and thus,
\begin{equation}
R_{2}(x)=\frac{\pi }{L_{0}}\sum_{m=1}^{\infty }P_{s_{m}}^{(0)}(x).  
\label{r2}
\end{equation}
It is important that all above distributions are closed expressions consistently describing the 
spectral characteristics in terms of periodic orbit theory.

\section{Spectral hierarchy}

As mentioned above, in general quantum graphs are not regular and so for them the spectral expansions 
of form (\ref{fn}) cannot be obtained directly. A generalization to the irregular case can be obtained 
by using the relationship between the two neighboring separator systems $\hat{k}_{n}^{(j)}$ and 
$\hat{k}_{n}^{(j-1)}$ and by applying Eq. (\ref{leveldelta}) to $f(k)=k$ at the $(j-1)$th level of the 
hierarchy:
\begin{equation}
\hat{k}_{n}^{(j-1)}=\int_{\hat{k}_{n-1}^{(j)}}^{\hat{k}_{n}^{(j)}}kdN^{(j-1)}.  
\label{kjint}
\end{equation}
Here, $N^{(j)}(k)$  corresponds to the spectral staircase of the sequence $\hat{k}_{n}^{(j)}$. 
Bootstrapping of the sequences $\hat{k}_{n}^{(j-1)}$ by  $\hat{k}_{n}^{(j)}$  (or $N^{(j-1)}(k)$  
by $N^{(j)}(k)$, see Fig. 1) means that $N^{(j-1)}\left( \hat{k}_{n}^{(j)}\right) =n$. Substituting 
expansion (\ref{gutzw}) for $N^{(j-1)}\left( \hat{k}_{n}^{(j)}\right)$ into Eq. (\ref{kjint}), 
and using $\hat{k}_{n}^{(j)}$ in the form
\begin{equation}
\hat{k}_{n}^{(j)}=\frac{\pi}{L_{0}}\left( n+\delta _{n}^{(j)}\right) ,
\label{knjdecomposition}
\end{equation}
we obtain the oscillating part of $\hat{k}_{n}^{(j-1)}$  in the form
\begin{equation}
\delta _{n}^{(j-1)}= f_{\delta }^{(j-1)}
-\sum_{p}C_{p}^{(j-1)}\sin \left(\omega _{p}^{(j-1)}n+\varphi _{p}^{(j-1)}\right) ,
\label{jfluct}
\end{equation}
Here, the zeroth term
\begin{equation}
f_{\delta }^{(j-1)} =\frac{1}{2}
\left(\delta _{n}^{(j)}-\delta _{n-1}^{(j)}\right) -\frac{1}{2}\left(
(\delta _{n}^{(j)})^{2}-(\delta _{n-1}^{(j)})^{2}\right) , 
\end{equation}
the amplitudes,
\begin{equation}
C_{p}^{(j-1)}=\frac{2}{L_{0}}\frac{A_{p}^{(j-1)}}{\omega^{(j-1)}_{p}}
\sin \frac{\omega^{(j-1)}_{p}}{2}\allowbreak \left(\delta _{n}^{(j)}-\delta
_{n-1}^{(j)}+1\right) , 
\end{equation}
and phases $\varphi_{p}^{(j-1)}=\omega^{(j-1)}_{p}\left(\delta _{n}^{(j)}+\delta _{n-1}^{(j)}-1\right)/2$ 
for every level $j$ are functions of the fluctuations $\delta _{n}^{(j)}$ and $\delta _{n-1}^{(j)}$ at the 
preceding hierarchy level.

Similar expansions are easily obtained for other spectral characteristics, for example, for 
$s_{n,m}^{(j-1)}=\hat k_{n+m}^{(j-1)}-\hat k_{n}^{(j-1)}$:
\begin{equation}
s_{n,m}^{(j-1)}=f^{(j-1)}_{s}+\frac{2}{L_{0}}\sum_{p}D_{p,m}^{(j-1)}\cos \omega^{(j-1)}_{p}
\left( n-\frac{m}{2}\varphi^{(j-1)}_{p}\right),
\label{snj}
\end{equation}
with the zeroth term
\begin{eqnarray}
f^{(j-1)}_{s}=s_{n,m}^{(j)}+\left(s_{n,m}^{(j)}-s_{n,m-1}^{(j)}\right) \times
\cr
\left(\pi m/L_{0}-( s_{n,m}^{(j)}+s_{n-1,m}^{(j)})/2 \right) 
-\xi_{n}^{(j)}\left( s_{n,m}^{(j)}-s_{n-1,m}^{(j)}\right),
\end{eqnarray}
where $\xi _{n}^{(j)}\,=(\delta _{n}^{(j)}+\delta _{n-1}^{(j)})/2$ and the expansion coefficients 
$\tilde D_{p,m}^{(j-1)}$ are obtained from the corresponding expansion for $s_{n,m}^{(j)}$. The 
equations relating the neighboring sequences can also be considered as describing the transition 
of a single separating sequence $f_{n}^{(j)}$ from one hierarchy level to another.

\section{Statistical description of spectral hierarchy}

As in the case of the regular graphs, the description of the stochastic properties of sequences 
such as $\delta _{n}^{(j)}$ or $s_{n,m}^{(j)}$ is based on the observation that parsing through 
the indices $n$ in the arguments of harmonic functions (\ref{jfluct}) and (\ref{snj})  leads to 
the appearance of random variables $x$. The idea of finding the distribution functions for various 
spectral characteristics is based on using the structural relations between the separating sequences 
obtained above in order to relate the probability distributions $P_{f}^{(j)}$ at different hierarchy 
levels. Beginning with the distribution $P_{f}^{(r)}$ at the regular level, one can determine the 
distribution $P_{f}^{(r-1)}$ at the next level and so on, ending with the last, physical level. 
\begin{figure}
\begin{center}
\includegraphics{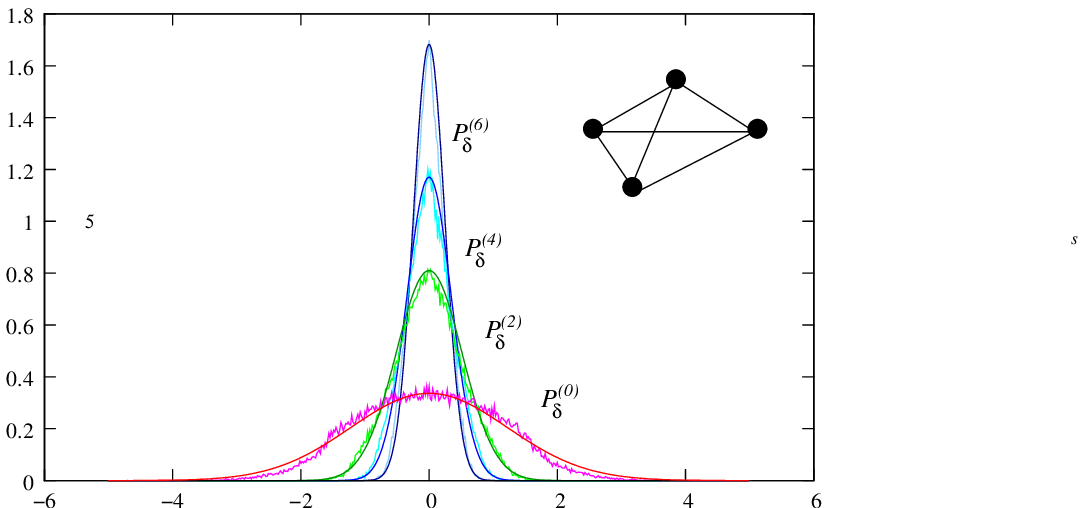}
\end{center}
\caption{Distribution of variances at the odd levels of the
spectral hierarchy for the four-vertex quantum graph with
$r=7$. The solid lines are the Gaussian approximations of the
numerically calculated histograms.}
\label{Fig.2}
\end{figure}
As an example, let us consider the behavior of the sequences $\delta _{n}^{(j)}$.  For simplicity, 
we treat the fluctuations $\delta _{n}^{(j)}$  and $\delta _{n-1}^{(j)}$ as independent random 
variables $\delta _{1}$  and $\delta_{2}$ distributed according to $P_{\delta }^{(j)}$.  
Correspondingly, one can write for the density $P_{\delta}^{(j-1)}(\delta)$
\begin{equation}
P_{\delta }^{(j-1)}=\int \delta \left(\delta
-\delta^{(j-1)}_{x}\right)
P_{\delta_{1}}^{(j)} P_{\delta_{2}}^{(j)}d\delta _{1}d\delta _{2}dx.
\end{equation}
Using Eq. (\ref{jfluct}) and representing the delta functional in exponential form, we obtain
\begin{equation}
P_{\delta }^{(j-1)}(\delta) =\int dke^{ik\delta }\left\langle
\prod_{p}\Lambda _{p}^{(j-1)}\left( x,\delta _{1},\delta
_{2}\right) dx\right\rangle _{\Omega^{(j-1)}},
\label{pdeltaj}
\end{equation}
Here, the factors $\Lambda _{p}^{(j)}\left( x,\delta _{1},\delta_{2}\right)$ correspond to the 
terms of expansion (\ref{jfluct}),  which are now explicit functions of fluctuations at preceding 
hierarchy levels, and $\left\langle \ast \right\rangle _{\Omega ^{(j)}}$ denotes averaging over 
these fluctuations with the weight
\begin{equation}
\Omega^{(j-1)}\left(\delta _{1},\delta _{2},k\right) =e^{-ikf_{\delta}^{(j-1)}
\left( \delta _{1},\delta _{2}\right) }P_{\delta}^{(j)}\left( \delta _{1}\right) 
P_{\delta }^{(j)}\left( \delta _{2}\right).
\end{equation}
The expression (\ref{pdeltaj})  generalizes regular expansions (\ref{pf}), 
(\ref{prime}) and (\ref{distreg}) for the single-level hierarchy to the general expressions for 
$j>0$, averaged over the disorder at the preceding levels. 
\begin{figure}
\begin{center}
\includegraphics{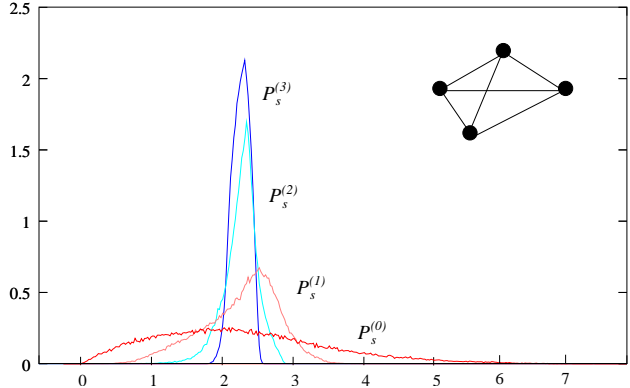}
\end{center}
\caption{Development of the probability distributions for the
distances between the nearest neighbors $s_{n}^{(j)}=\hat{k}_{n}^{(j)}-\hat{k}_{n-1}^{(j)}$, $r=3$.
The maximum distance between the nearest neighbors at the $j = 0$  level in this case is $s_{max}=8.68$, 
for the regular cell size $\pi/S_{0}= 2.28$.}
\label{Fig.3} 
\end{figure}
We note that the argumentation concerning the Gaussian distribution form in Section 2 
\cite{Proxorov,Revesz} can be directly applied to the distribution of $\delta_{x}^{(r)}$  
at the regular level. However, as shown in Fig. 2, the distribution of $\delta^{(j)}_{x}$ 
at higher levels $j>0$ is also Gaussian-like. For other spectral characteristics, for 
example, $s_{n}^{(j)}$  (see Fig. 3), the sequence of transitions of form 
(\ref{pdeltaj}) can lead to asymmetric (non-Gaussian) distributions.

\section{Discussion}

The method proposed in \cite{Anima}  for solving the spectral problem is based on establishing 
the structural relationships between the sequence of physical levels $k_{n}$ and the regular 
sequence $\hat{k}_{n}^{(r+1) }$ specified as an explicit function $\hat{k}_{n}^{(r+1)}=\hat{k}^{(r+1)}(n)$. 
For quantum graphs, the regular sequence is given by (\ref{mid}) and relation to $k_{n}$ is 
established through the system of auxiliary sequences $\hat{k}_{n}^{(j)}$, bootstrapping $k_{n}$ 
with  $\hat{k}^{(r+1) }(n)$. The spectral hierarchy thus obtained consists of the system of 
sequences $\hat{k}_{n}^{(j)}$  and transition equations (\ref{kjint}) from 
$\hat{k}_{n}^{(j)}$ to $\hat{k}_{n}^{(j-1)}$.

This approach allows not only the description of the evolution of base sequences $\hat{k}_{n}^{(j)}$ 
from low to high hierarchy levels, but also the complete probability description of spectral 
characteristics in the framework of periodic orbit theory including those that are not directly 
described by the Gutzwiller formula. In this case, it is possible to follow the development of 
the scales of spectral fluctuations, distributing disorder over the intermediate hierarchy levels, 
gradually passing from less to more disordered sequences. While the base sequence is maximally 
ordered, the amplitude of fluctuations in each next sequence $\hat{k}_{n}^{(j)}$ increases as 
the index $j$  decreases, i.e. with the approach to the physical spectrum \cite{Anima}. The 
minimum number of auxiliary sequences $\hat{k}_{n}^{(j)}$  necessary for bootstrapping $\hat{k}_{n}^{(r+1)}$ 
with $\hat{k}_{n}^{(0)}$ defines to the complexity of the spectral problem with respect to the 
given bootstrapping method.

The above relation between the properties of the series of expansions (\ref{jfluct}) and the 
properties of weakly dependent random variables \cite{Proxorov,Revesz} reveals the physical 
origins of the universality of the distributions of different spectral characteristics following 
from the limiting properties of the sums of such quantities.  
The existence of a sufficient number of transitions between hierarchy levels of irregular systems 
and, correspondingly, of averaging processes over random phases and disordered sequences 
$\hat{k}_{n}^{(j)}$ in Eq. (\ref{pdeltaj}) leads not only to the Gaussian shape of the distribution 
of probabilities $P_{f}^{(0)}$ [as, e.g., for $\delta N(k)$ and, correspondingly for $\delta _{n}^{(0)}$, 
see \cite{ABS}  and Fig. 2), but also to the appearance of more complex (e.g., Wignerian, see \cite{BGS} 
and Fig. 3) distributions.

It is also important that determining the fluctuation probabilities in form (\ref{pdeltaj}) 
makes it possible not only to follow the appearance of general, universal statistical relations, 
but also to describe in detail the specific features of distributions $P_{f}^{(j)}$, which present 
the individual properties of each particular system.

\bigskip

Work supported in part by the Sloan--Swartz Foundation.

\end{document}